\documentclass[conference]{IEEEtran}
\IEEEoverridecommandlockouts
\usepackage{cite}
\usepackage{amsmath,amssymb,amsfonts}
\usepackage{algorithmic}
\usepackage{graphicx}
\usepackage{textcomp}
\usepackage{xcolor}

\usepackage{multirow}
\usepackage{booktabs}
\usepackage{subfigure}
\usepackage{tikz}
\usetikzlibrary{positioning, shapes.geometric}

\usepackage{tikz,xcolor,hyperref}

\def\BibTeX{{\rm B\kern-.05em{\sc i\kern-.025em b}\kern-.08em
    T\kern-.1667em\lower.7ex\hbox{E}\kern-.125emX}}
\begin{document}

\title{A Framework for Health-informed RUL-constrained Optimal Power Flow with Li-ion Batteries}

\author{Jiahang Xie, Yu Weng, and Hung D. Nguyen
\thanks{Authors are with School of EEE, NTU, Singapore. \textit{jiahang001, yu006@e.ntu.edu.sg, hunghtd@ntu.edu.sg.}}}

\maketitle

\begin{abstract}
Battery energy storage systems are widely adopted in grid-connected applications to mitigate the impact of intermittent renewable generations and enhance power system resiliency. Degradation of the battery during its service time is one of the major concerns in the deployment that strongly affects the long-term lifetime. Apart from environmental factors, this intrinsic property of a battery depends on the daily operating conditions. Thus, optimally engaging the daily operation of the battery based on its current status in order to meet the required remaining useful life becomes a practical and demanding need. To address this issue, this paper proposes a health-informed RUL-constrained optimal power flow framework to characterize the corresponding optimal feasible operation space. The targeted service lifespan is achieved if the battery's working condition is confined within this feasible domain. Equivalent box constraints are then constructed for better computational efficiency in solving the optimization problem. In this framework, a Monte Carlo-based data-driven approach and a health indicator (HI) representing the battery's current states are introduced. The performance of the proposed method is illustrated with the IEEE 39-bus system.

\end{abstract}

\section{Introduction}

Battery energy storage systems (BESS) find a wide range of applications in power systems owing to its dispatchability, interruptibility, and energy efficiency \cite{divya2009battery,atwa2009optimal}. Following the surging integration of renewable generations, electric vehicles, and data centers, optimal operation with BESS needs to be designed to both alleviate intermittency and maximize the economic benefits \cite{barton2004energy,gayme2012optimal}. 


In the literature, OPF with BESS has been studied extensively \cite{divya2009battery,dall2013distributed}. Many existing works consider the time scale of a day \cite{gayme2012optimal,chandy2010simple}. Also, the battery is assumed to work under fixed mode, defined by the maximum current and voltage, through out their lifetime regardless of the fact the battery condition changes over time. This assumption may lead to inappropriate decisions for economic operation and management of the battery. Due to an increasing BESS deployment in power systems, the need for optimally operating batteries on a daily basis, relying on their health states to prolong their remaining useful life (RUL), becomes more practical and demanding.

RUL is defined as the duration the battery serves from the current state to the maximum equivalent full cycles \cite{RUL,lifetime_characteristics,specifications}. RUL can be estimated based on the battery health condition. RUL information plays an important role in deciding the short-term operation and regulations to achieve an expected lifetime. Unfortunately, estimating RUL is typically challenging because the uncertainty in the long-term operation is difficult to quantify and keep track as the time passes \cite{sankararaman2013remaining}. Being a fundamental tool for operation, the OPF framework rarely considers years-time scale as well as the RUL requirement which is complicated in nature. This work therefore investigates how operational conditions affect RUL and how the battery health state should be quantified. The gained insights will help incorporating RUL into OPF problem \cite{review}.

To address RUL-constrained OPF problem with the knowledge on the battery health condition, a Monte Carlo (MC)-based data-driven framework is proposed to learn the nonlinear relationships among RUL constraint and operational constraints. The feasible operational regions with which the RUL target can be achieved will be established. Health indicator (HI) is brought up as a single index to represent the battery's health state instead of using various internal variables of the cells \cite{resistance}. From the feasible operational regions and HI, inner-approximated box subspaces are constructed to offer great computation efficiency in finding the optimal operating condition. The main contributions of this paper are:
\begin{itemize}
    \item Propose a data-driven framework to learn the dependence of RUL on the battery's operational constraints. 
    \item Form the foundation of health-informed RUL-constrained OPF (HIRUL-OPF) problem by converting RUL constraint into equivalent operational constraints. 
    \item Construct box constraints for a given RUL target to make the RUL-constrained OPF more tractable. Reveal the pattern of how the box region varies with HI values under the same RUL requirement. 
    
    
\end{itemize}

\section{RUL-constrained OPF} 
The OPF problem with an RUL constraint of the battery will be presented, wherein the current operating conditions are linked up with a prediction index that contains RUL information on a long time scale.
\subsection{Optimal Power Flow with Batteries}\label{sec:opf_Battery}
A typical OPF problem can be formulated as \cite{nguyen2018constructing, HungGPOPF}:
\begin{equation}\label{eq:opf}
\begin{aligned} 
    \min_\textbf{u} \quad & J(\textbf{x}, \textbf{u}) \\
    \text{subject to} \quad & \textbf{g}(\textbf{x}, \textbf{u}) = 0; \,\, \textbf{h}(\textbf{x}, \textbf{u}) \leq 0.
\end{aligned}
\end{equation}
The objective is to minimize the cost function $J$ that may consist of the cost of the power generation and power losses. The controllable vector $\textbf{u}$ contains independent decision variables, such as active power injections from generators and batteries \cite{chandy2010simple}. Vector $\textbf{x}$ denotes the state variables including reactive power injections from PV buses, PQ buses' nodal voltages, and transferred power flows. Equality constraints $\textbf{g}(\cdot) = 0$ can be power flow equations. Inequality constraints may represent operational limits such as generators' maximum capacity, voltage constraints, and thermal limits. 

The conventional OPF \eqref{eq:opf} above is extended to incorporate batteries. Particularly, equality constraints will include the batteries' power injections. Inequality constraints also comprise the following operational constraints of battery \cite{levron2013optimal}:
\begin{align} \label{eq:IneC_B}
  \textbf{V}^{min}_{Brate} \leq & \textbf{V}_b \leq \textbf{V}^{max}_{Brate}, \\ \label{eq:IneC_B2}
   0 \leq & \textbf{E}_b \leq   \textbf{E}_{Brate},
  \\ \label{eq:IneC_B3}
 \textbf{P}_{Bch}  \leq  &  \textbf{P}_b  \leq  \textbf{P}_{Bdisc}.
\end{align}
Here $\textbf{V}_b$ is the controllable vector denoting battery operating voltage. $\textbf{V}^{max}_{Brate}$ and  $\textbf{V}^{min}_{Brate}$ are the voltage lower and upper bounds. $\textbf{E}_b$ is the stored energy inside battery, which can be described by state of charge (SOC), while $\textbf{E}_{Brate}$ is the maximum battery storage capacity. $\textbf{P}_b$ is the battery discharging/charging power which is confined within $ \textbf{P}_{Bch}$ and $ \textbf{P}_{Bdisc}$.


\subsection{Battery RUL Constraint}
\label{sec:opf_RUL}

Degradation of the batteries is an inevitable process regardless of whether the batteries are used or not. Remaining useful life or RUL is an important and handy index indicating how long the battery performs decently. RUL thus naturally associates with the incurred operation cost and the value asset of the battery. For a newly deployed energy storage, a common practice is to set the so-called ``expected'' useful life, for example, 3-5 years for Li-ion batteries. However, the actual lifetime of a battery is strongly affected by the practical operation conditions, that unfortunately are different from the standard conditions used by the manufacturers \cite{RUL}. Misuse of batteries might shorten their lifetime, so the expected useful life will not be met. We, therefore, consider the RUL requirement into the OPF framework to optimize the short-term operation of the battery that not only minimizes the economical costs but also guarantees the long-term performance expressed in terms of RUL. The new constraint is the following:
\begin{equation}
    T \leq RUL,
    \label{eq:RUL_Csts}
\end{equation}
wherein $T$ is the expected time threshold and $RUL$ is the estimated duration till when the battery functions properly.

However, the RUL-constraint \eqref{eq:RUL_Csts} cannot be applied to OPF problem \eqref{eq:opf} directly as it is an indicator manifested in the long term while OPF chiefly concerns the current operation. In connection to this, RUL often requires a rigorous estimation over a long course of the operation period, maybe up to several years. A data-driven framework is thus proposed to handle the RUL constraint in the OPF settings, by converting \eqref{eq:RUL_Csts} into operational constraints applicable to OPF. 
It is worth mentioning that, in the literature, the lifetime information has also been incorporated into the daily operation of devices, typically using the operation cost \cite{park2009optimal}. This approach might be handy as the operation cost of the battery can limit the inappropriate usage of the batteries. However, the operation cost also needs to be estimated carefully to avoid too rough estimation based on the provided lifetime from the manufactures as well as to reflect the degradation process. This work does not focus on such an operational cost, but the feasible operating range of the batteries that, from an engineering view, directly links to the degradation and the practical operation. 

\section{RUL-equivalent operational constraints}\label{sec:RUL-equivalent operational constraints}
In this section, RUL-equivalent operational constraints will be constructed for HIRUL-OPF. We propose a data-driven framework using Monte Carlo method to estimate respective feasible operational domains for the battery. 

\subsection{MC-based Data-driven Framework}\label{sec:Data-driven Framework}

\begin{figure}[b]
\vskip -2em
\centerline{\includegraphics[scale=0.5]{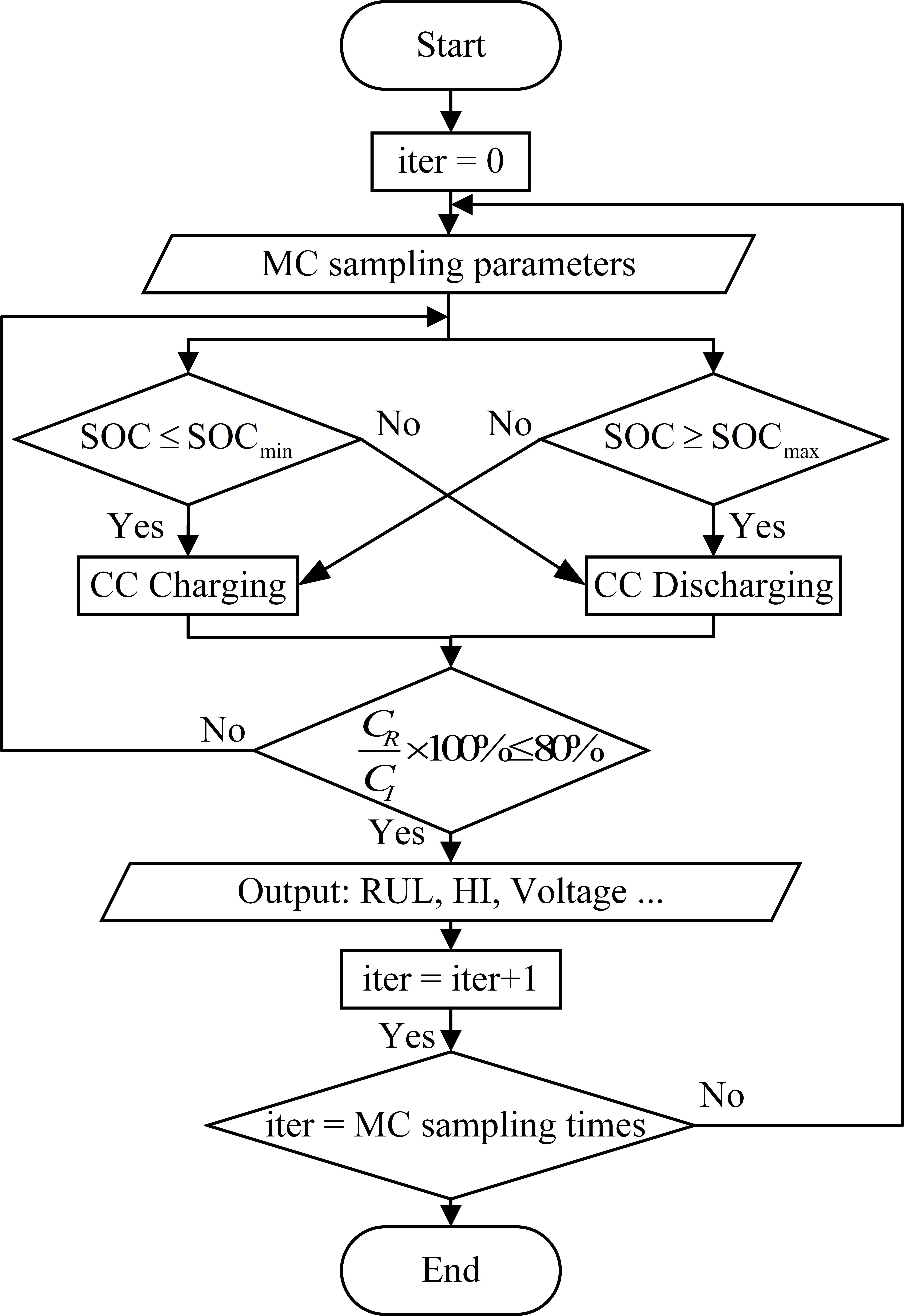}}
\caption{Data-driven framework flow chart}
\vskip -2em
\label{fig:MC-Data-driven flow chart}
\end{figure}

This MC-based approach is proposed to estimate the RUL of a battery if the battery follows strictly operational constraints defined by the operating charge/discharge currents and allowed voltage ranges. The basic flow of this framework is shown in Fig.~\ref{fig:MC-Data-driven flow chart}. When the repeated sampling has not reached the setting times, random values will be given to the chosen inputs, and each sample represents one scenario under a specified battery state or operation conditions. In each scenario, with the initial inputs, run the battery simulation continuously until satisfying the termination condition, from which the RUL information can be obtained. The termination condition for each simulation is $\frac{C_R}{C_I}\times 100\% \leq 80\% $ \cite{review}, which denotes the battery reaches the end-of-life when the remaining capacity $C_R$ is less than $80\%$ of the initial capacity $C_I$.

From \eqref{eq:IneC_B}-\eqref{eq:IneC_B3}, important constraints of battery are bounds on voltage $\mathbf{V}^{max}_{Brate}$, $\mathbf{V}^{min}_{Brate}$, energy $\mathbf{E}^{max}_{Brate}$ and power $\mathbf{P}_{Bdisc}, \mathbf{P}_{B\,ch}$. For a RUL target, the data-driven method will find the corresponding voltage bounds $\mathbf{V}^{max}_{B\text{rate-RUL}}, \mathbf{V}^{min}_{B\text{rate-RUL}}$, energy bound $\mathbf{E}^{max}_{B\text{rate-RUL}}$ and  power bounds $\mathbf{P}_{Bdisc-RUL}, \mathbf{P}_{Bch-RUL}$ used in HIRUL-OPF.

The energy bound $\textbf{E}^{max}_{B\text{rate-RUL}}$ is limited by keeping $SOC_{\text{max}}$ and $SOC_{\text{min}}$ within certain ranges following uniform distribution. The voltage upper bound $\textbf{V}^{max}_{B\text{rate-RUL}}$ is the voltage when SOC reaches the maximum value, and the lower bound $\textbf{V}^{min}_{B\text{rate-RUL}}$ is the battery cut-off voltage. The power bounds $\textbf{P}_{Bdisc-RUL}, \textbf{P}_{Bch-RUL}$ then can be computed using the charging/discharging current $\textbf{I}_{Bch-RUL}, \textbf{I}_{Bdisc-RUL}$ in the constant current operation mode. 



\subsection{Relationship between RUL and Voltage Bounds}\label{sec:Relationship between RUL and Voltage Bounds}
\begin{figure}[t]
\vskip -2em
\centerline{\includegraphics[scale=0.5]{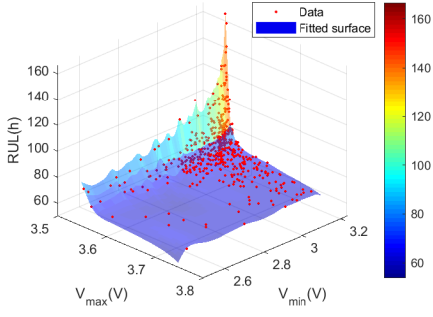}}
\caption{The relationship of RUL and voltage bounds}
\label{fig:rul_vmax_vmin}
\vskip -1em
\end{figure}

Below are simulation settings for creating the data for RUL and voltage bounds $\textbf{V}^{max}_{B\text{rate-RUL}}, \textbf{V}^{min}_{B\text{rate-RUL}}$. The battery chosen in this work is $LiFePO_4$ cell \cite{specifications}. Its expected maximum lifespan is 1000 equivalent full charge/discharge cycles while the initial full cycle is 500 when the simulation starts. The charging/discharging current are 4.3A and 11.7A with 500 MC samples, $SOC_{\text{max}}$ in $60\%-100\%$, and $SOC_{\text{min}}$ in $0\%-40\%$. The output data are RUL and battery voltage $\textbf{V}^{max}_{B\text{rate-RUL}}, \textbf{V}^{min}_{B\text{rate-RUL}}$, denoted by $\textbf{V}_{max}, \textbf{V}_{min}$ in Fig.~\ref{fig:rul_vmax_vmin}. 

The RUL-voltages data acquired from MC method are denoted by the red points in Fig.~\ref{fig:rul_vmax_vmin}. The RUL ranges between 60 and 160 hours when $\textbf{V}_{max}$,$\textbf{V}_{min}$ fall in the intervals of [3.5V, 3.8V] and [2.4V, 3.3V]. Based on the raw data, the fitted surface depicting voltage bounds and RUL is demonstrated with color map, indicating the trend of how RUL changes when the battery operates with different voltage bounds. Smaller $\textbf{V}_{max}$ and larger $\textbf{V}_{min}$ lead to a longer RUL.

To further verify the trend and significance of the above relationships, correlation coefficients and p-value are calculated and shown in Table~\ref{tab:Correlation coefficients and p-value with RUL}. Pearson correlation coefficient of two random variables is a measure of their linear dependence as defined in \eqref{eq:pearson}. p-value describes significance of the relationship between factors. Smaller p-value indicates larger statistical significance.
\begin{equation}
    \rho_{X,Y}=\frac{\sum_{i=1}^n(x_i-\overline{x})(y_i-\overline{y})}{\sqrt{\sum_{i=1}^n(x_i-\overline{x})^2}\sqrt{\sum_{i=1}^n(y_i-\overline{y})^2}}
    \label{eq:pearson}
\end{equation}

\begin{table}[t]
\centering
\caption{Correlation coefficients and p-value with RUL under constant charging/discharging current}
\begin{tabular}{ccc}
\toprule
         & Correlation coefficient & p-value$^\mathrm{a}$ \\ \midrule
$\textbf{V}_{max}$  & -0.5097                 & 1.9e-34 \\
$\textbf{V}_{min}$  & 0.2856                  & 7.7e-11 \\
$\Delta\textbf{V}$ & -0.4108                 & 8.8e-22 \\ \bottomrule
\end{tabular}
\vskip -2em
\label{tab:Correlation coefficients and p-value with RUL}
\end{table}
In Table \ref{tab:Correlation coefficients and p-value with RUL}, p-value$^\mathrm{a}$ denotes value at 0.05 confidence level. The voltage upper bound and operational range are negatively correlated with RUL, which is due to larger $\Delta\textbf{V}$ and $\textbf{V}_{max}$ indicate larger SOC range and mean value, i.e. deeper charging/discarging process, which are detrimental for battery health, thus resulting in shorter RUL. Whereas $\textbf{V}_{min}$ is positively correlated with RUL as its increment leads to smaller SOC range. However, the p-value with respect to $\textbf{V}_{min}$ is larger, which means less significant since higher $\textbf{V}_{min}$ not only reduces SOC range but also raises SOC mean value. This property also makes its correlation coefficient smaller. Thus, voltage upper bound $\textbf{V}_{max}$ is selected for further investigation as it has higher significant factor for RUL with smaller p-value. The numerical results prove consistent with the trend in Fig.~\ref{fig:rul_vmax_vmin}.

\subsection{Relationship between RUL, Current and Voltage upper bound}\label{sec:Relationship between RUL, CD Current and vmax}

\begin{figure}[b]
\vskip -2em
\centering
\subfigure[The relationship of RUL, $\textbf{V}_{max}$ and $I_{discharge}$]{
\begin{minipage}[t]{0.49\linewidth}
\centering
\includegraphics[scale=0.47]{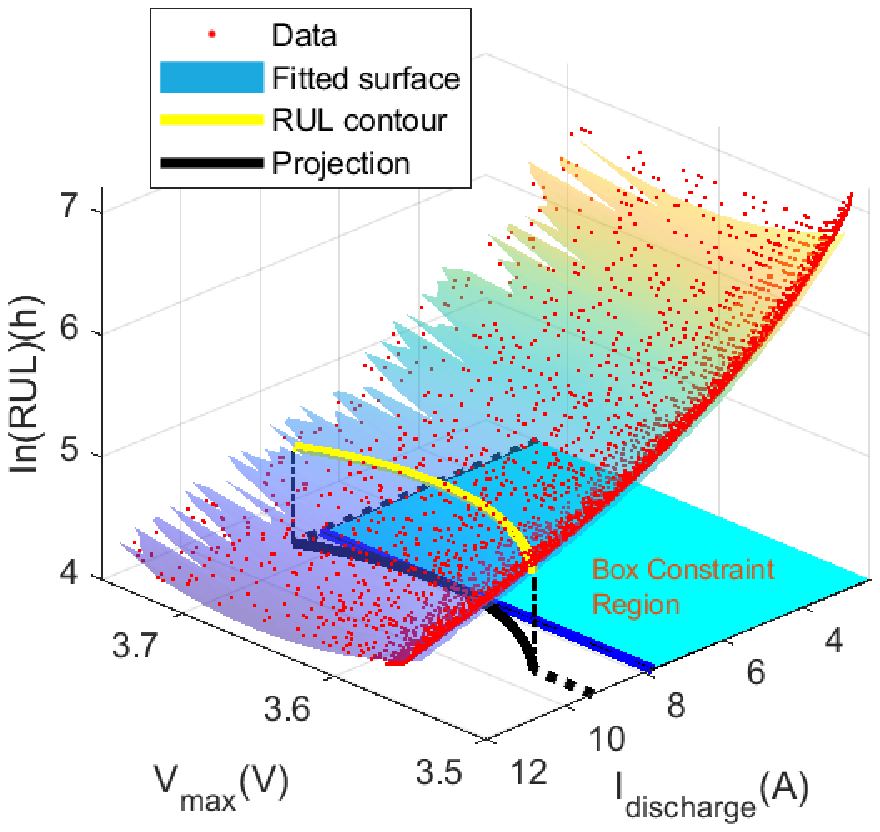}
\end{minipage}%
}%
\subfigure[The relationship of RUL, $\textbf{V}_{max}$ and $I_{charge}$]{
\begin{minipage}[t]{0.49\linewidth}
\centering
\includegraphics[scale=0.47]{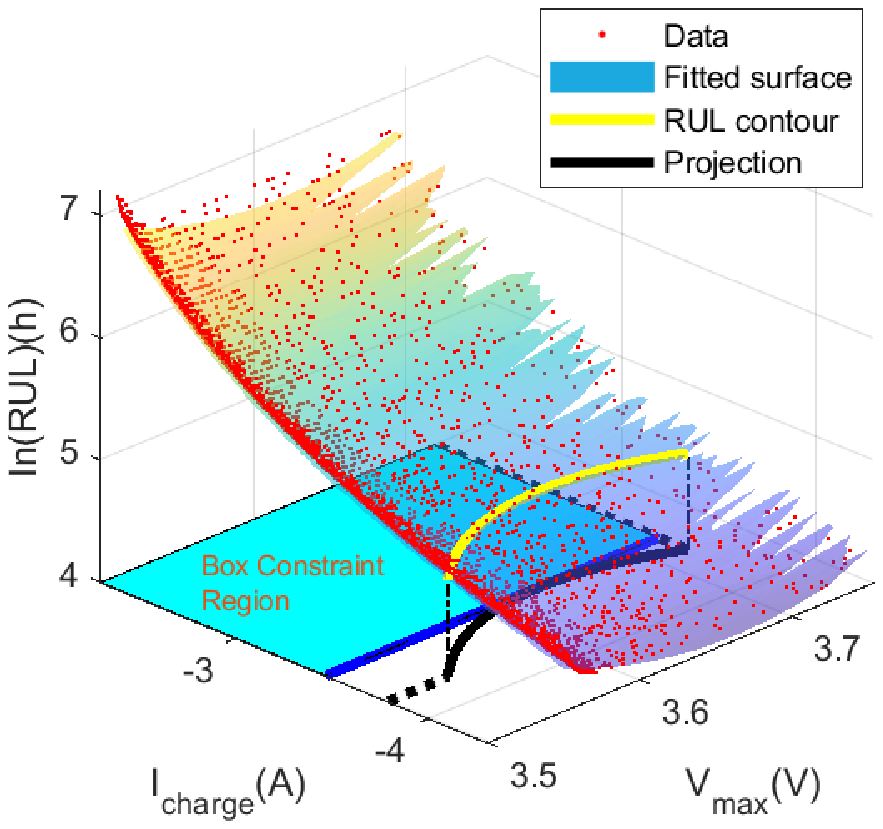}
\end{minipage}%
}%
\centering
\caption{The relationship of RUL, $\textbf{V}_{max}$ and charging/discharing current}
\label{fig:rul_vmax_Icharge_Idischarge}
\end{figure}

From the above section, voltage upper bound $\textbf{V}_{max}$ is selected to reflect RUL. Here $\textbf{V}_{max}$ together with charging/discharging current will be further analyzed in the relationship with RUL. The number of MC samples is 10000 with $SOC_{\text{min}}=0\%$, $SOC_{\text{max}}$ in $60\%-100\%$, charging/discharging current in 1C-2C and 1C-5C. The output data are RUL and battery voltage $\textbf{V}^{max}_{B\text{rate-RUL}}$, denoted by $\textbf{V}_{max}$ in Fig.~\ref{fig:rul_vmax_Icharge_Idischarge}.

The acquired data and derived relationship among RUL, $\textbf{V}_{max}$ and charging/discharging current $\textbf{I}_{charge}$ $\textbf{I}_{discharge}$ is shown in Fig.~\ref{fig:rul_vmax_Icharge_Idischarge}. After taking the logarithm of RUL to narrow the range of its variation for an explicit trend illustration,  $ln(\text{RUL})$ stays in [4h,7h] with $\textbf{V}_{max}$ ranges in [3.5V,3.8V]. As depicted in Fig.~\ref{fig:rul_vmax_Icharge_Idischarge}, smaller charging/discharging current and smaller voltage upper bound will result in longer RUL.

\begin{figure}[t]
\vskip -1em
\centerline{\includegraphics[scale=0.4]{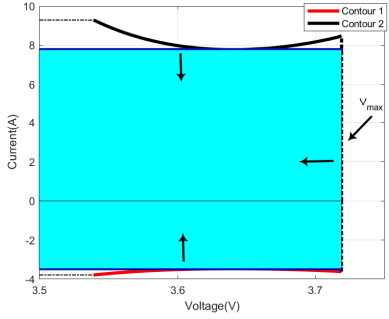}}
\caption{A box constraint feasible region}
\label{fig:region}
\vskip -2em
\end{figure}

To achieve RUL target, the feasible operating domain is confined in RUL contour-limited region in Fig.~\ref{fig:rul_vmax_Icharge_Idischarge}, with RUL-constraint contour in yellow. When operating voltage and current vary within this region, RUL can be kept above the yellow contour, i.e., battery can serve longer than the required RUL. Moreover, since the resulting feasible region is non-convex, the region can be convexified using its inner-approximated box region highlighted in bright blue, ``sacrificing'' only current boundaries without interfering voltage range. The box region is shown in Fig.~\ref{fig:region}. The two contours are the RUL contour projected to $\textbf{I}_{charge}, \textbf{I}_{discharge}$ in a 2-D space from Fig.~\ref{fig:rul_vmax_Icharge_Idischarge}.

The box region specifies the possible operating voltage and current bounds for a given RUL target. $\textbf{V}_{max}$ is the dotted line while maximum current is the dark blue line. When RUL is expected to achieve a higher level, i.e., ~RUL contour with larger Z-axis value in Fig.~\ref{fig:rul_vmax_Icharge_Idischarge}, box constraint region will shrink in the direction indicated by the arrows in Fig.~\ref{fig:region}.

Learning relationships between RUL and voltage/current bounds, the recommended operational feasible domain has been found, as the approximated box region shown in Fig.~\ref{fig:region}, with which the converted RUL constraints are applicable to OPF problem. The following section shows further analysis on how the battery current states affect the corresponding operational constraints for the same RUL requirement. 

\section{HI and operational constraints} \label{sec:HI and operational constraints}

As discussed earlier, the main objective of this work is to transform the RUL constraint into operational constraints that are more computationally efficient. At a specific point in the service time, such operational constraints depend on the current status of the battery or how healthy the battery is. In general, the feasible operation domain is large for a new battery and shrinks gradually when it ages. This section focuses on the Health Indicator (HI), which is a concise index to represent the battery's health status. The relationship between battery health conditions and operation constraints will be investigated and then used to estimate the recommended operating conditions for ensuring the RUL objective.


\subsection{Definition of HI}
HI should be chosen from the variables that can represent the battery health degradation. It can either be a battery inherent parameter, like battery internal resistance and maximum capacity, or other measured index such as full charging/discharging time and average voltage falloff in equal time interval, etc. In this paper, internal resistance \cite{resistance} is chosen as the HI as it is not restricted to certain types of charging/discharging protocols. Also it has the advantage that the data can be directly exported from the lithium-ion battery module within the whole charging/discharging process. So the normalized HI is defined as below:

\begin{equation}
    HI=\frac{R_{EOL}-R_X}{R_{EOL}-R_{BOL}} 
    \label{eq:normalization}
\end{equation}

In which $R_{EOL}$ is the internal resistance at the end-of-life point corresponding with $HI=0$. $R_{BOL}$ is the internal resistance at the beginning-of-life with defined $HI=1$. $R_X$ denotes the current internal resistance. With equation \eqref{eq:normalization}, the HI is normalized and mapped into the section [0,1], with which the current battery stage can be represented.

\subsection{Relationship between HI and Operation Constraints}
\begin{figure}[t]
\vskip -2em
\centering
\subfigure[The relationship of HI and $\textbf{V}_{width}$]{
\begin{minipage}[t]{0.49\linewidth}
\centering
\includegraphics[scale=0.4]{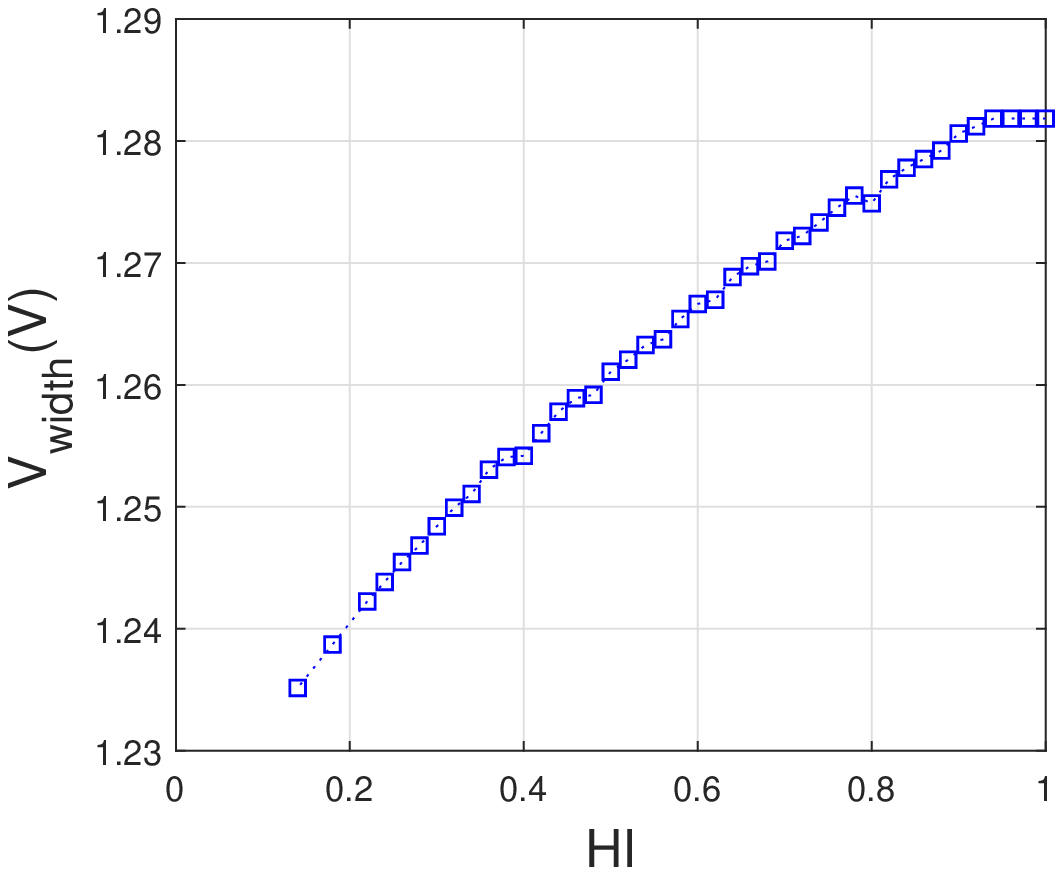}
\end{minipage}%
}%
\subfigure[The relationship of HI and $\textbf{I}_{length}$]{
\begin{minipage}[t]{0.49\linewidth}
\centering
\includegraphics[scale=0.4]{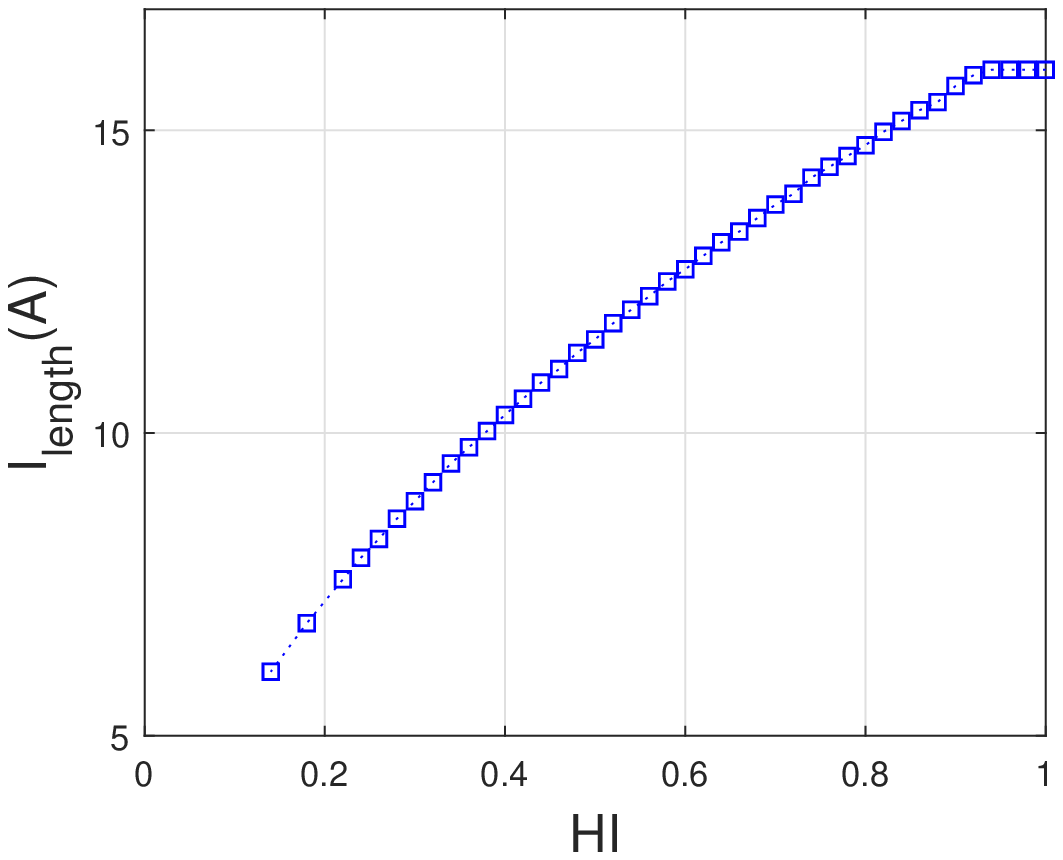}
\end{minipage}%
}%
\centering
\caption{The relationships among HI, $\textbf{V}_{width}$, and $\textbf{I}_{length}$}
\vskip -2em
\label{fig:hi_Vrange_Irange}
\end{figure}

To learn how the operational constraints change with HI under the same RUL target, the data-driven framework in section \ref{sec:Data-driven Framework} is used for data acquiring and subsequent analysis. $SOC_{\text{min}}$ is set as $0\%$ while $SOC_{\text{max}}$ varies from $90\%$ to $100\%$. The initial full cycles of the battery are randomly sampled from $0$ to $1000$ corresponding to brand new and the end of service. The number of MC samples is 100. Under each scenario, the box constraint region and operational boundaries are constructed for each HI value using the binary search algorithm. Taking the maximum $\textbf{I}_{charge}, \textbf{I}_{discharge}$ and $\textbf{V}_{max}$ as the critical points from the rectangular area in Fig.~\ref{fig:rul_vmax_Icharge_Idischarge} and Fig.~\ref{fig:region}, through which the size of the box region is the output, i.e., $\textbf{I}_{length}$ and $\textbf{V}_{width}$ in Fig.~\ref{fig:hi_Vrange_Irange}.

From the acquired data and for a given RUL target, how the box constraint region shrinks following the battery's deterioration can be analyzed. As shown in Fig.~\ref{fig:hi_Vrange_Irange}, the ranges of voltage and current in the box region decrease when HI value drops. When HI is close to 1, the ranges of both current and voltage can reach the maximum values regarding specific value of $T$, as shown by the line segment parallel to the HI axis. Based on Fig.~\ref{fig:region} and Fig.~\ref{fig:hi_Vrange_Irange}, the suggested operational constraints can be established for the expected RUL with a given battery current health state. 


\section{Numerical Results and Discussions}\label{Test Results}
The proposed method is verified on IEEE 39-bus system \cite{39_bus} using MATPOWER 7.0 for OPF. Case 1 is the conventional OPF problem without RUL constraint, while Case 2 is HIRUL-OPF for battery. Each battery contains 1,500 $LiFePO_4$ cells in series with 11.4 kWh capacity in total \cite{cell_number}. Bus 36, 37 and 38 are set as nodes with battery, corresponding to the 7-$th$, 8-$th$ and 9-$th$ generation sources. The initial equivalent full cycles of the 3 batteries are 100, 500, and 700 to indicate their current health state. Their respective HI are 0.9, 0.5, and 0.3. In Case 2, the RUL target $T$ is 120 hr. 

\begin{figure}[ht]
\vskip -2em
\centerline{\includegraphics[scale=0.25]{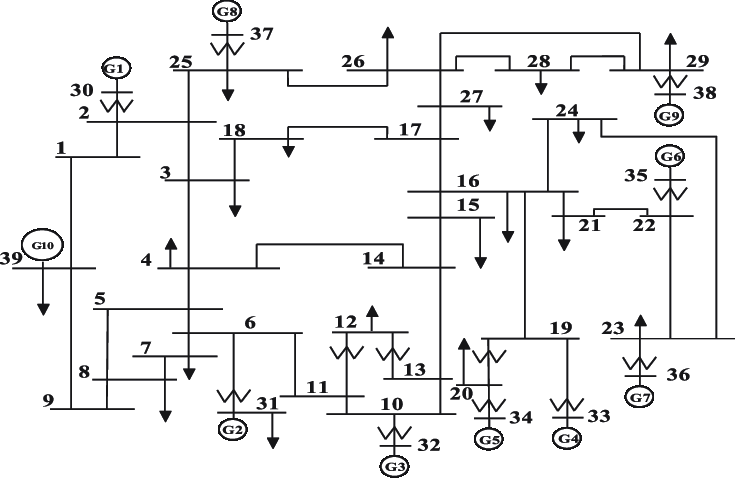}}
\caption{39-bus testing system}
\label{fig:39_bus}
\end{figure}

\begin{figure}[t]
\centering
\vskip -1.8em
\subfigure[Bus voltage]{
\begin{minipage}[t]{0.49\linewidth}
\centering
\includegraphics[scale=0.45]{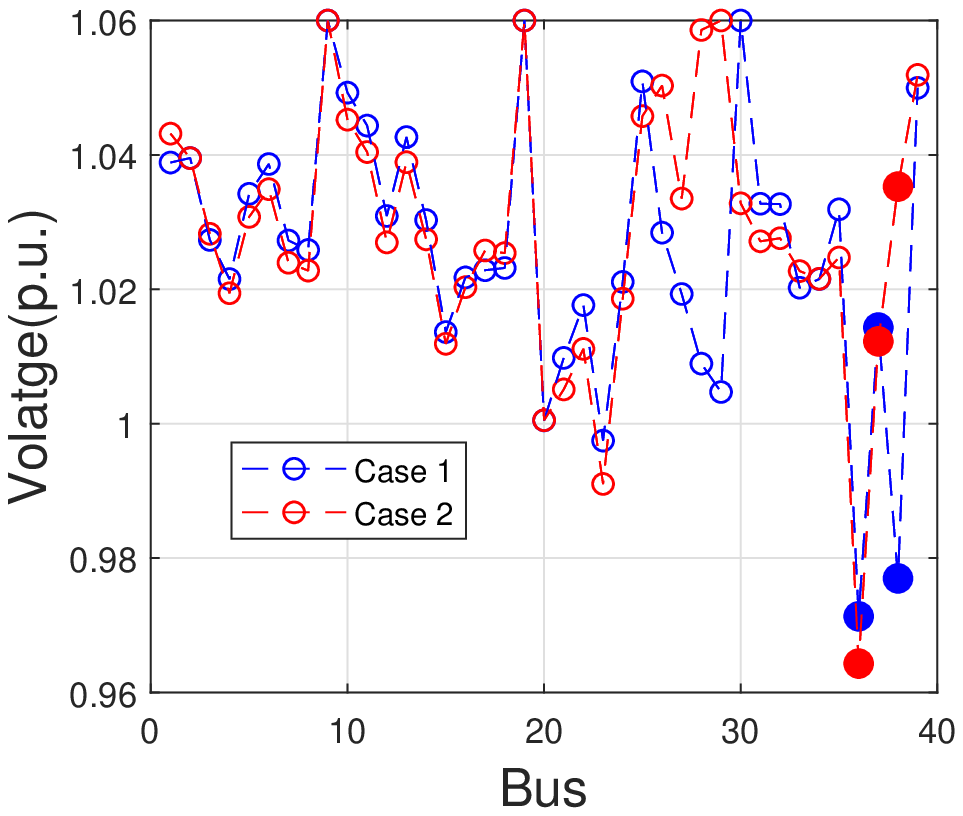}
\end{minipage}
}%
\subfigure[Active power generation]{
\begin{minipage}[t]{0.49\linewidth}
\centering
\includegraphics[scale=0.45]{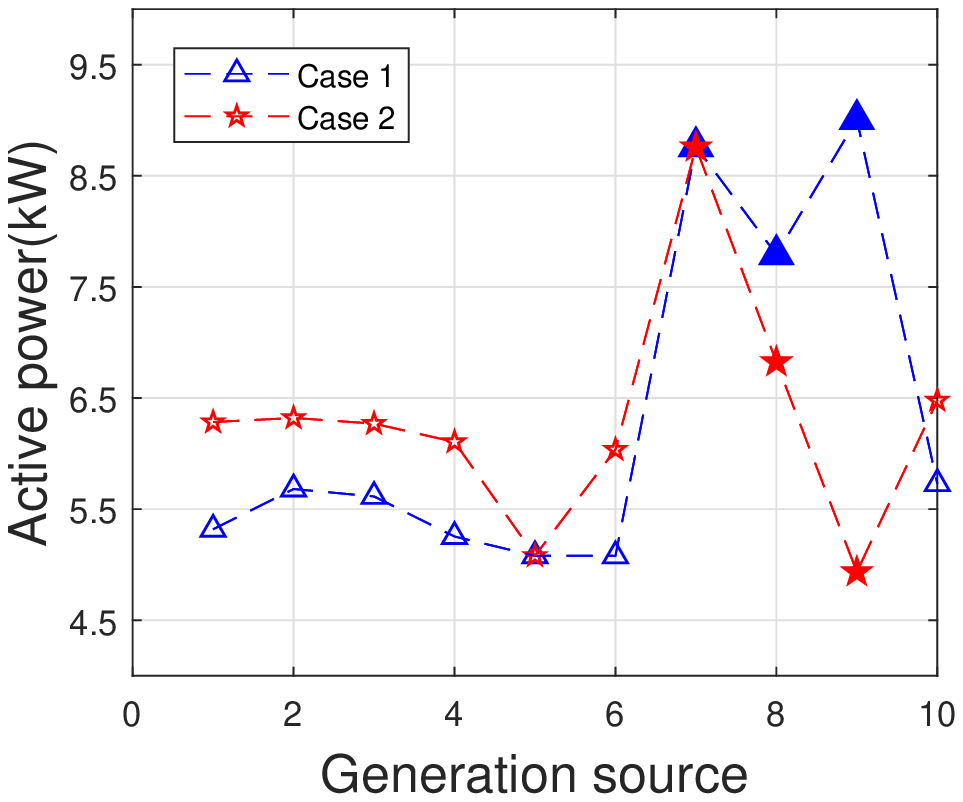}
\label{fig:39-bus OPF solutions (b)}
\end{minipage}%
}%
\centering
\caption{OPF solutions of the 39-bus system in two scenarios}
\vskip -2em
\label{fig:39-bus OPF solutions}
\end{figure}

The voltage and power results are plotted in Fig.~\ref{fig:39-bus OPF solutions}, wherein the filled dots are battery buses. Fig.~\ref{fig:39-bus OPF solutions} shows bus voltages in [0.96p.u.,1.06p.u.] and active power in [4.5kW, 9.5kW]. The voltage and power change due to the RUL constraint. In the beginning, the 3-$rd$ battery deteriorates the most, so its power has been limited to the narrowest range. Thus, a drop in generation is detected in Fig.~\ref{fig:39-bus OPF solutions (b)}, with output power increment of other generators, showing load pick-up behavior.


Table~\ref{tab:opf_results} compares the power, voltage, and RUL of batteries in Case 1 and Case 2. The powers are smaller in Case 2 with RUL constraint, especially on bus 38, with a max of $4.9344$ kW about half of that in Case 1. Table~\ref{tab:opf_results} shows the node voltages maintain the same level on Case 1 and Case 2, which indicates OPF performance is stable whereas the costs increase due to the added RUL constraint. The RUL of batteries has a large improvement in Case 2 and satisfies the target. This shows the effectiveness of the proposed approach.

In Case 2, comparing the values of nodes 36-38, the power at bus 38 is limited the most since the battery degrades severest. To reach the required 120h RUL, its power limit is tighter than other batteries. This shrinking tendency is consistent with the  analysis in Fig.~\ref{fig:hi_Vrange_Irange}, which shows the confined ranges of voltage and current get smaller gradually (power is the product of voltage and current) from Bus 36 to 38 following the decreasing HI from $0.9$ to $0.3$. 

\begin{table}[b]
\vskip -2em
\caption{OPF Solutions In 2 Cases}
\begin{center}
\begin{tabular}{c|c|cc}
\toprule
\multicolumn{2}{c|}{Quantities}                & Case 1 & Case 2 \\ \hline
\multicolumn{2}{c|}{Cost (\$/hr)}              & 29949  & 32972  \\ \hline
\multirow{3}{*}{BESS Power (kW)}     & Bus 36 & 8.7640 & 8.7572 \\ \cline{2-4} 
                                     & Bus 37 & 7.7917 & 6.8238 \\ \cline{2-4} 
                                     & Bus 38 & 9.0079 & 4.9344 \\ \hline
\multirow{3}{*}{\begin{tabular}[c]{@{}c@{}}BESS Voltage (p.u.)\\ (With $\textbf{V}^{max}_{B\text{rate-RUL}}$ in Case 2)\end{tabular}} & Bus 36 & 0.9713 & 0.9643 (1.0599) \\ \cline{2-4} 
                                     & Bus 37 & 1.0143 & 1.0122 (1.0581) \\ \cline{2-4} 
                                     & Bus 38 & 0.9770 & 1.0353 (1.0569) \\ \hline
\multirow{3}{*}{RUL (hr)}            & Bus 36 & 115.7  & 120.1  \\ \cline{2-4} 
                                     & Bus 37 & 61.22  & 120.7  \\ \cline{2-4} 
                                     & Bus 38 & 36.07  & 120.6  \\ \bottomrule
\end{tabular}
\label{tab:opf_results}
\end{center}
\vskip -2em
\end{table}


\section{Conclusion}
This paper focuses on a novel approach for estimating the feasible operational regions that the battery needs to comply with to meet the RUL requirement, based on the current health status denoted concisely by the Health Indicator. The estimated feasibility domains will be used in the proposed health-informed RUL-constrained OPF to design the optimal daily operation of the battery. The relationships among the battery's RUL, HI, and operational constraints are studied based on the proposed data-driven framework. By leveraging these relationships, the RUL constraint is converted into battery operational constraints, most importantly, box constraints in charging/discharging current and maximum voltage. Also, the effects of HI on the scale of the box constraint region are also analyzed. Further research includes modeling of the battery's partial cycle in optimization while considering the time dependence under different HI and environmental factors.

\section*{Acknowledgment}
This work is supported by NTU SUG, MOE Tier-1 2019-T1-001-119, EMA \& NRF EMA-EP004-EKJGC-0003.

\bibliographystyle{IEEEtran}
\bibliography{main.bbl}

\end{document}